\documentclass[doublespacing]{elsart}
\usepackage[english]{babel}
\usepackage{bm}

\usepackage{epsfig}
\usepackage{graphics}

\usepackage{amssymb}

\begin{document}

\begin{frontmatter}

\title{Can electron beams be really focused by bent crystals?}

\author{Victor V. Tikhomirov}
\ead{vvtikh@mail.ru}
\address{Research Institute for Nuclear Problems, Bobruiskaya str., 11, 220050, Minsk, Belarus.}%


\begin{abstract}
The short period atomic plane modulation is suggested to be
applied to modify the electron channeling potential in order to
make it possible to considerably increase electron channeling
efficiency by the crystal structure brake ensured by either a
plane crystal cut or amorphous layer beneath the crystal entrance
surface. The achieved channeling efficiency of 60-70 \% can
considerably facilitate the process of high energy electron beam
focusing by specially cut bent crystals. Possible parameters of
one TeV electron beam focusing region are estimated using
simulations
\end{abstract}



\end{frontmatter}
\section{Introduction}
The possibility to deflect channeling particles by bent crystals
is known since 1976 [1]. Its applications for both positively
charged particle beam extraction and collimation have been widely
demonstrated and possess perspectives to be used at the LCH and
FCC [2-4]. Another promising application of bent crystals is beam
focusing [5-8]. Since the effective field strength of crystal
planes exceeds one kilotesla [9], bent crystals are able to focus
high energy beams within a centimeter focal length. The high
efficiency of all the bent crystal applications to positively
charged particle beam manipulation originates from both the high
channeling stability and capture probability, the latter of which
can be additionally increased by a crystal structure break [10,
11].

However besides positron beam sharp focusing, future $e^+ e^-$
colliders will also need the same of negatively charged electron
ones. Meanwhile, experiments demonstrate [12, 13] that electron
beam deflection efficiency remains relatively small even for thin,
moderately bent crystals. Besides the strong electron scattering
by nuclei, the poor electron channeling effeciency originates also
from the unsuitable electron planar potential coordinate
dependence, which also makes the method [10, 11] of channeling
efficiency increase practically inapplicable.

Developing the advantages of the electron dechanneling rate fall
at TeV energies [14], we suggest in this Letter to modify the
electron planar potential by a short period small amplitude atomic
plane modulation [15, 16] in order to further decrease the
electron dechanneling rate as well as to increase, both directly
and by the method [10, 11], the probability of electron capture
into the stable channeling motion. We expose justified parameters
of the focusing region to clarify the perspectives of bent crystal
application at both the ILC and CLIC as well as to provide a
benchmark for possible focusing scheme improvements.


\section{Channeling efficiency increase by both short period
plane modulation and crystal structure break}\label{s2}

\begin{figure}
\label{Fig1}
 \begin{center}
\resizebox{90mm}{!}{\includegraphics{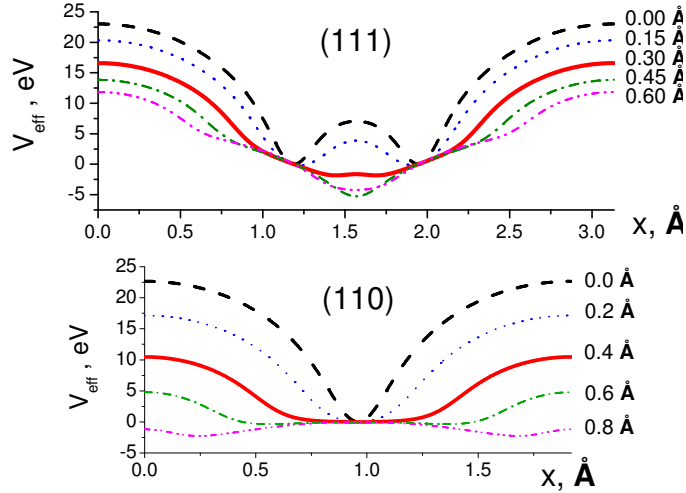}}\\
\caption{Averaged potential of (111) (top) and (110) (bottom) Si
planes, modulated with the amplitudes, indicated on the right, and
averaged over the modulation period.}
\end{center}
\end{figure}

A salient feature of the electron average potential (i. e. of the
dependence of electron potential energy in the averaged field of
crystal planes on the transverse coordinate $x$, measured along
the normal to the latter) are sharp dips (one for (110) and two
for (111) transverse lattice period) in the region of high nuclear
density - see Fig. 1, resulting both in the small acceptance of
the low transverse energy states and enhanced nuclear scattering,
immediately depopulating the latter [14]. In its turn the region
of the highest transverse energies of channeled electrons both
readily shrinks at a moderate crystal bending and is intensively
depopulated by the strong transverse energy fluctuations
$\delta\varepsilon_\bot = \varepsilon v_x \theta_s$, induced by
the electron nuclear scattering [14, 17] at a random angle
$\theta_s$ and enhanced by the large electron channeling
oscillation velocity $v_x \simeq \theta_{ch}$ within the high
nuclear density region, where $\theta_{ch}$ is critical channeling
angle.

\begin{figure}
\label{Fig2}
 \begin{center}
\resizebox{75mm}{!}{\includegraphics{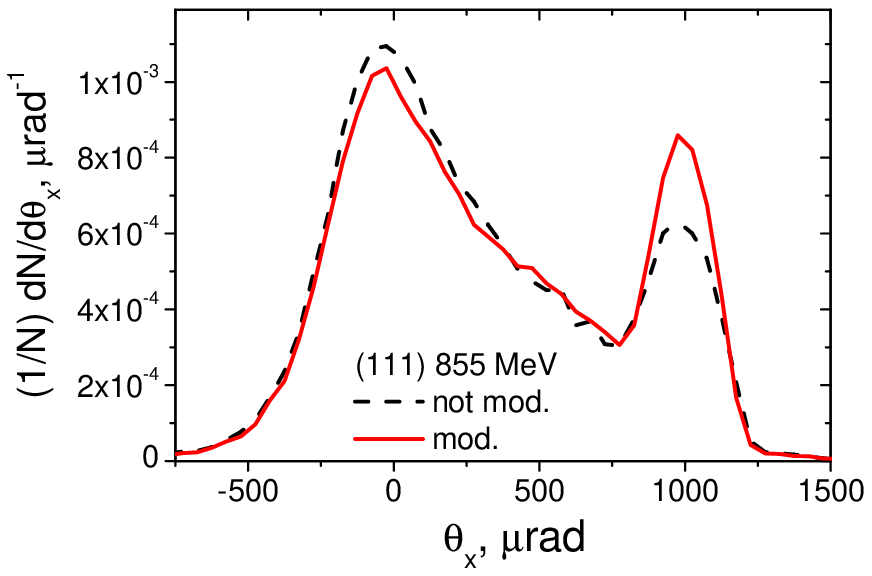}}\\
\vspace{5 mm}
\resizebox{75mm}{!}{\includegraphics{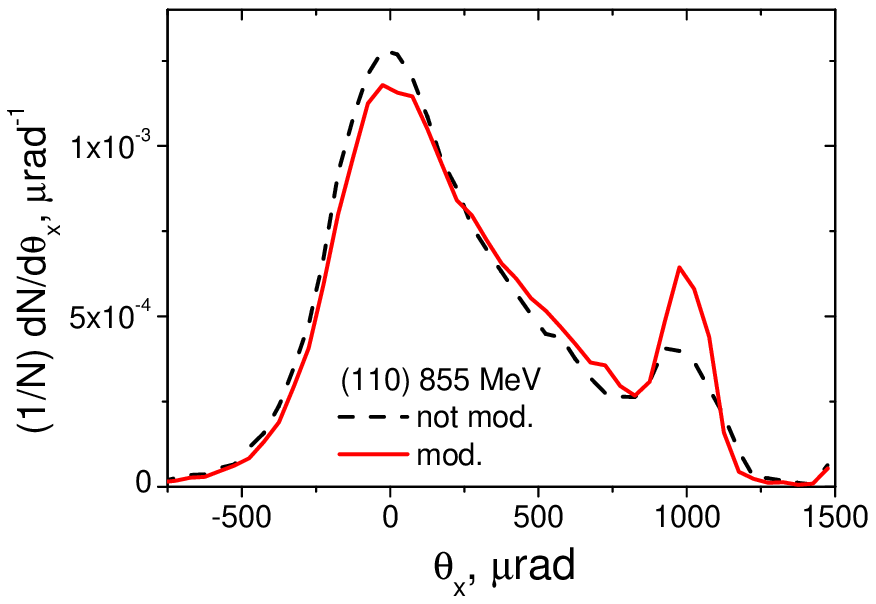}}\\
\caption{Angular distributions of 855 MeV electrons after passing
through a 30 $\mu$m Si crystal at the incidence on (111) (top) and
(110) (bottom) crystal planes with zero angle and 20 $\mu$rad
divergence.}
\end{center}
\end{figure}
All these circumstances explain the low, about 20\% in Fig. 2,
efficiency of channeled 855 MeV electron deflection, observed in
the MAMI experiment [12] in a 30 $\mu m$ Si (111) crystal, well
reproduced by the simulation method [18, 12]. Fig. 2 also
demonstrates the better deflection efficiency of (111) planes in
the case of electrons. Note also that only (111) channeling plane
orientation is available in ultrathin crystals [12]. Besides the
mentioned above direct factors favoring the fast electron
dechanneling, a large dispersion of the channeling electron
oscillation period makes unfeasible the method of channeling
efficiency increase by a crystal structure break [10, 11]. Owing
to this reasoning one can readily assume that both the fraction
value and stability of electron channeling can be increased by
smearing out the potential dips to about a half of the transverse
lattice period.

Putting aside an evident but technically too challenging
possibility of a considerable thermal vibration amplitude increase
by an intense crystal heating, we prefer the recently suggested
[15] and already realized [16] idea of the small amplitude short
period crystal undulator. Such a Si crystal undulator with
modulated (periodically bent) (110) planes was grown by the method
of molecular beam epitaxy. The (110) planes modulation appeared
[16, 19] due to the periodic introduction of Ge atoms into the Si
crystal growing in the $\langle 100 \rangle$ direction. If the
modulation (undulator) period $\lambda_u$, is much shorter than
the period $\lambda_{ch}$ of channeling motion (see Fig. 1 in
[15]), the latter is governed by the planar potential additionally
averaged over the modulation period $\lambda_u$. This "second
averaging" naturally smears our the unwanted sharp potential dips
over the range determined by the modulation amplitude $a$. Fig. 1
demonstrates that the potentials of the modulated (111) and (110)
planes well enough approach harmonic ones at a considerable part
of the channel width $d$ at modulation amplitudes $a$ = 0.3 $\AA$
and $a$ = 0.4 $\AA$, respectively. Fig. 2 demonstrates that the
modification of both planar potentials by the modulation results
in the dechanneling process deceleration at $250$ $\mu rad <
\theta_x < 750$ $\mu rad$ as well as in the 15-20\% relative
increase of the deflection efficiency at $\theta_x > 800$ $\mu
rad$. However, since both these effects turn out to be moderate,
to reach a really decisive increase of the negatively charged
particle channeling efficiency we suggest here to apply the idea
of crystal structure break introduction [10, 11] becoming feasible
for electrons in crystals with the short period plane modulation.
\begin{figure}
\label{Fig3}
\hspace{1 cm} \resizebox{120mm}{!}{\includegraphics{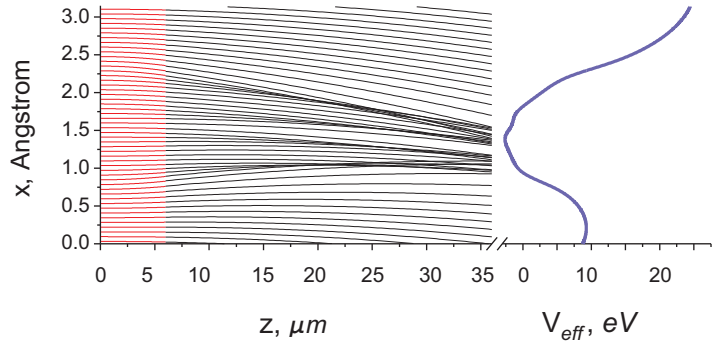}}\\
 \begin{center}
\caption{One TeV electron focusing to the most stable channeling
region inside a bent modulated Si (111) crystal channel. }
\end{center}
\end{figure}
\begin{figure}
\label{Fig4}
 \begin{center}
\resizebox{80mm}{!}{\includegraphics{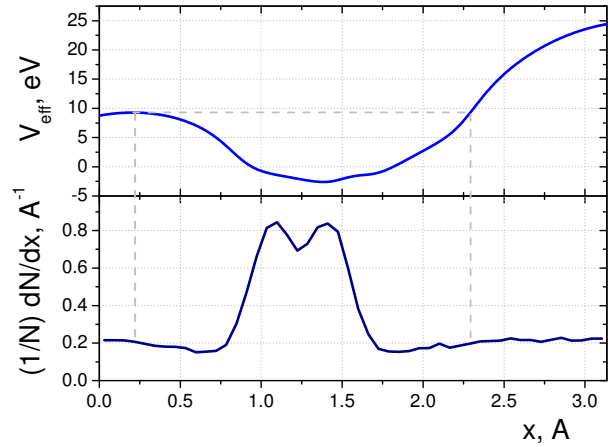}}\\
\caption{Effective electron potential of the bent modulated Si
(111) planes (top) and transverse coordinate electron distribution
(bottom) within one (111) transverse period $d$. }
\end{center}
\end{figure}

Channeling efficiency can be increased by a slight focusing of
highly parallel particle beam within each channel. Following [10,
11] such a focusing can be accomplished by a thin ($0.08
\lambda_{ch} - 0.12 \lambda_{ch}$) crystalline plate separated
from the crystal bulk by either an empty [10] or amorphous [11]
plane layer of the thickness $0.12 \lambda_{ch} - 0.17
\lambda_{ch}$. Since both of the approaches [10, 11] preserve a
single crystal structure both in the front and behind its break,
each channel of the front thin crystal plate can serve as a lens
for a one in the crystal bulk. Note also, that possible
deformations accompanying both the break formation and crystal
bending can be compensated by an appropriate choice of particle
beam incidence direction to the planes (of about one microradian
in the one TeV example below). The discussed approach works well
in a semi-parabolic focusing potential, approached by the natural
inter-planar potential for positively charged particles and, as
suggested above, also by the planar potential of a short-period
modulated crystal for negatively charged ones (see Fig. 1).
Another necessary condition of a very low incident beam divergence
is for sure fulfilled for the quite small emittance beams of the
future $e^+e^-$ linear colliders (see Table 30.1 in [20]).

Fig. 3 illustrates by fifty trajectories the channeling efficiency
increase of one TeV electrons in a bent Si (111) crystal.
Electrons are first accelerated towards the channel center by the
focusing planar field at $z < 6$ $\mu m$ (red color on the left),
and then (see both Figs. 3 and 4) freely shift towards the
potential minimum inside the crystal structure break region of $6$
$\mu m < z < 36$ $\mu m$. The effective potential of bent (111)
planes is shown both on the right in Fig. 3 and on the top in Fig.
4.

\section{One TeV electron beam focusing region example}\label{s2}

Below we will proceed from the ideas of Refs. [5-7] for positively
charged beams to present a possible geometry of one TeV electron
focusing region which implements additionally the idea of crystal
structure break [10, 11]. It was realized recently [8] that
focusing crystals with a skew back face [6, 7] will encounter
technological problems at centimeter focal lengths. To avoid the
latter, "a bent plane-parallel silicon plate whose side edges are
rotated at a small angle with respect to crystallographic planes"
[8] was suggested. We will consider the crystal assisted focusing
of a one TeV electron beam with initial transverse size $\Sigma_x
= 1$ $mu m$ as an example. Fig. 5 depicts both the focusing
crystal parameters and electron trajectories in the focusing plane
$xz$, being also the plane of channeling oscillations and crystal
bending. It is assumed that electron focusing in the normal $yz$
plane is accomplished by the usual means.
\begin{figure}
\label{Fig5}
 \begin{center}
\resizebox{110mm}{!}{\includegraphics{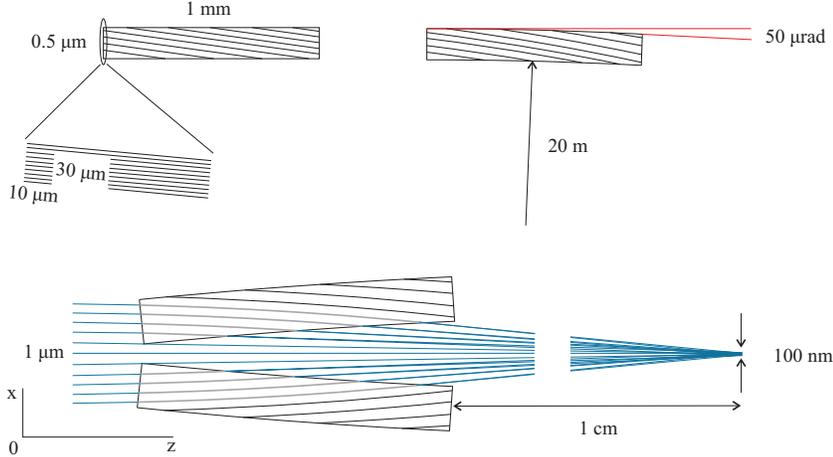}}\\
\caption{One TeV electron beam focusing. Bent crystal dimensions
and the crystal cut region (top, left), bending radius and angle
(top, right), beam focusing by a pair of crystals, focusing
length, focus size and electron trajectories in the focusing $xz$
plane (bottom).}
\end{center}
\end{figure}
Below we present the simulation results allowing us to demonstrate
the consequences of both the short period crystal modulation and
crystal structure brake as well to explain both the tight
interrelation and severe limitations of the dimensions of the
electron focusing region depicted in Fig. 5. Figs. 6 and 7
present, respectively, the differential and the integral electron
angular distributions behind the Si (111) and Si (110) crystals,
revealing that the crystal structure brake application to a
crystal with the short period plane modulation makes it possible
to increase the \textbf{electron deflection efficiency} up to the
appropriate levels of 70 and 60 \% for (111) and Si (110) planes,
respectively.

Rather close values of the channeling inefficiency at the crystal
entrance on the one hand and the dechanneling loss percentage
inside the crystal on the other, which both have dropped down to
15\% in Si (111) and 20 \% in Si (110), reflect the reached
optimal balance of these devastating processes. Note that electron
dechanneling at such a low loss level can't be described by a
single exponential function which only could make rigorous
dechanneling length introduction possible [14]. Though the rate of
dechanneling process considerably falls at one TeV, it still
limits the \textbf{crystal thickness} $l = 1$ $mm$ measured along
the electron velocity. The same do the radiative losses, which can
be simulated by the method [21-23].
\begin{figure}
\label{Fig6}
 \begin{center}
\resizebox{63mm}{!}{\includegraphics{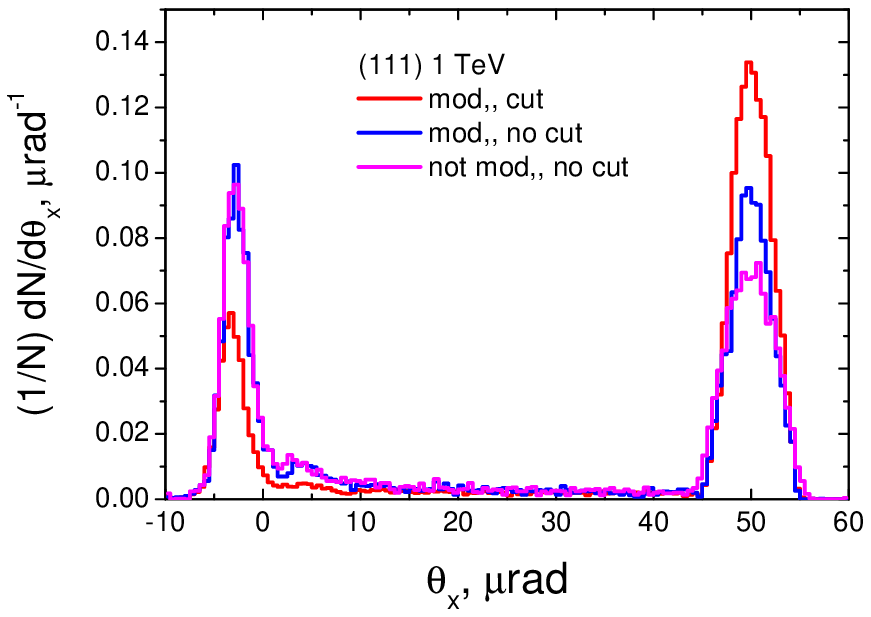}},\hspace{1 cm}
\resizebox{62mm}{!}{\includegraphics{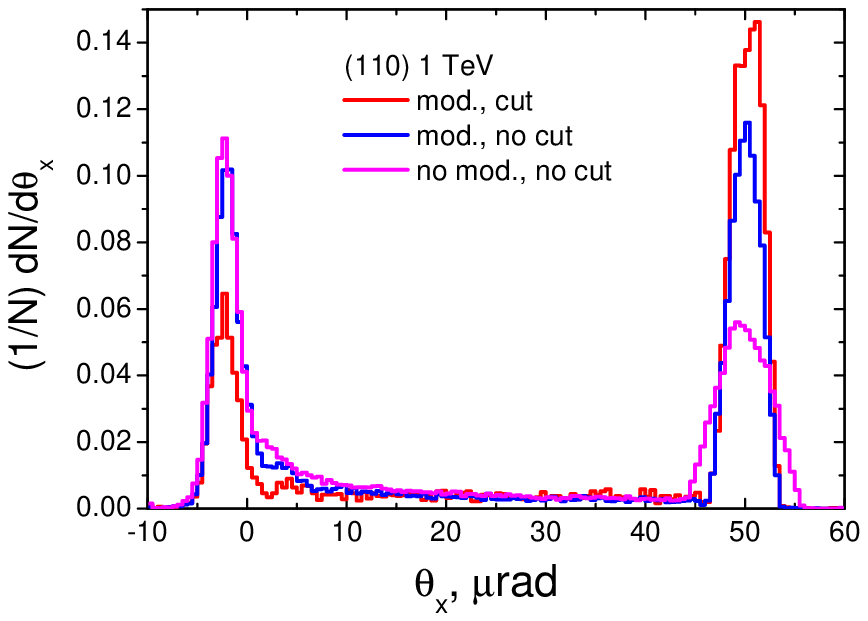}}\\
\caption{Electron angular distribution in the focusing plane for
unmodified Si (111) (left) and (110) planes (no mod., no cut), the
same for the planes, modulated with a short period and 0.3 \AA
~and 0.4 \AA ~amplitudes (mod., no cut) and for the latter with a
crystal structure brake extending from $z = 10$ $\mu m$ to $z =
40$ $\mu m$ (mod., cut). }
\end{center}
\end{figure}
\begin{figure}
\label{Fig7}
 \begin{center}
\resizebox{60mm}{!}{\includegraphics{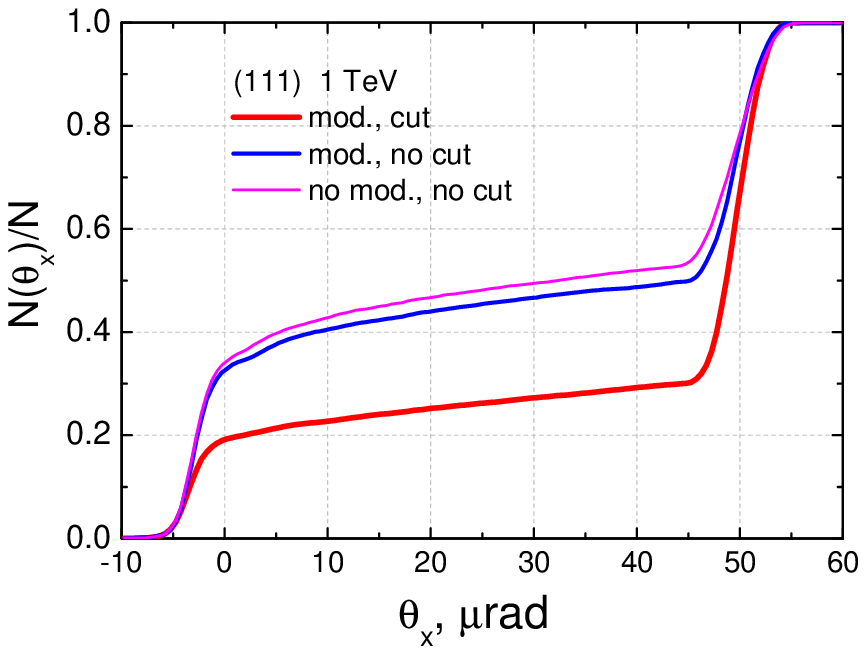}},\hspace{1 cm}
\resizebox{60mm}{!}{\includegraphics{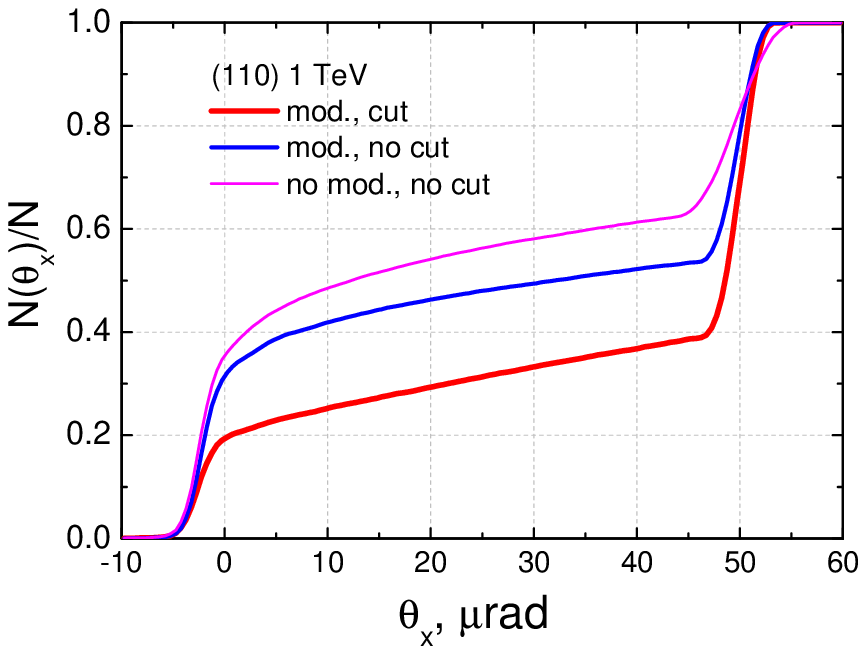}}\\
\caption{Integral distributions corresponding to the differential
ones from Fig. 6. }
\end{center}
\end{figure}

Crystal \textbf{bending radius} R = 20 cm amounts to about a dozen
of the minimal bending radii $R_{min} = \varepsilon/e/E_{max}$,
where $E_{max}$ is the maximal planar field. Note that the
practically used radii indeed usually exceed much the value
$R_{min}$, demonstrating that the later is not a completely
optimal characteristic. One could introduce instead an alternative
one of $R_{V_0} = \varepsilon d/V_0$, corresponding to the equal
to $V_0$ centripetal potential variation at the period $d$ of
planar potential, where $V_0$ is the amplitude of the latter.
Since $R_{V_0} \sim 10$ $m$ for both of the considered crystal
planes, one should realize that the choice of $R \approx 2
R_{V_0}$ is sufficiently grounded and can be additionally adjusted
through a thorough optimization only. Knowing now both the crystal
thickness and bending radius one can readily find the
\textbf{crystal bending angle} $\varphi = l/R = 50 \mu rad$.

Let us discuss the choice of \textbf{focal length} of $f = 1$
$cm$. Since channeled electron beam inevitably acquires an angular
divergence determined by the channeling angle $\theta_{ch}$ (of
about 5 $\mu rad$ at 1 TeV), one should surely limit the focal
length from above reducing thus the focus size $\sigma_x = f
\theta_{ch}$ in the focusing plane. One the other hand, since the
focus size has to be, at least, an order less than the beam size
$\Sigma_x \simeq f \varphi$ before focusing, the focal length has
been also limited from below, resulting in total in the compromise
choice of $f = 1$ $cm$.

The latter value determines the \textbf{focus sizes} in both $xz$,
$\sigma_x = f \theta_{ch} \sim 50$ $nm$, and $yz$, $\sigma_y = f
\theta_{ys} ~ 15$ $nm$, planes, where $\theta_{ys} \approx 1.53$
$\mu rad$ is the root mean square scattering angle in the $yz$
plane extracted from the simulated angular distribution depicted
in Fig. 8.
\begin{figure}
\label{Fig8}
 \begin{center}
\resizebox{60mm}{!}{\includegraphics{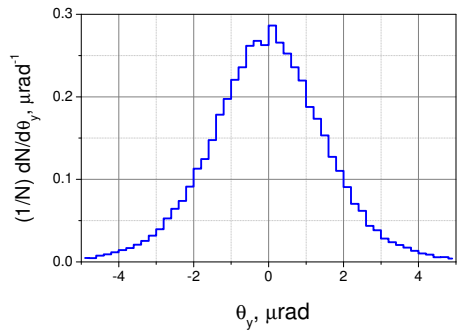}}\\
\caption{Electron angular distribution in the plane $yz$ normal to
the focusing plane for the Si (111) orientation. }
\end{center}
\end{figure}

The positron beam focusing region parameters can also be readily
estimated. Since well channeled positrons scatter on atomic
electrons only, their dechanneling length more then two orders
exceeds that of electrons [14]. In addition, due to the more
suitable averaged potential coordinate dependence, positrons are
both better captured into channeling and liable to the direct
application of the method [10, 11] of channeling efficiency
increase. The most pronounced consequence of these advantages is a
reduction of the positron multiple scattering angle in the $yz$
plane at least by an order of value. The \textbf{focus size} in
the same plane can be accordingly reduced to about one nanometer.

Since channeling induces the positron beam divergence of about the
channeling angle, as in the electron case, the positron beam focal
size in the focusing $xz$ plane can't be reduced so pronouncedly.
Nevertheless, provided a small enough incident beam divergence
[20], the method [10, 11] of crystal structure break can be
applied to optimize bending radius, percentage and transverse
energy spread of channeled positrons in order to diminish the
focal size down, say, to 20 nm by means of both focal length and
angular divergence reduction. Note that, alternatively, initial
transverse size of the focused positron beam can be increased by
an order or more at the cost of some focus size increase in both
$xz$ and $yz$ planes.

\section{Conclusions}

Both the short period atomic plane modulation and crystal
structure brake have been suggested to be applied to increase the
efficiency of electron deflection by bent crystals. It has been
demonstrated through simulations that the deflected electron
percentage of 60-70 \%, which justify enough the crystal
application for beam focusing, is achievable for 1 TeV electrons.
The estimates of focus sizes for both 1 TeV electron and positron
beams in the focusing $xz$ and normal $yz$ planes, evaluated for 1
mm focusing Si crystal bent with 20 m radius, are summarized in
the Table.

The latter reads that the intrinsic channeling angular spread
makes it hardly possible to reach a 10 nm focus size in the
focusing plane both for electrons and positrons. In its turn
nuclear scattering limits from below the possible focal size in
the normal plane by 10 nm for electrons, while the less intense
scattering of channeled positrons makes the same of 1 nm
achievable. Since the critical bending radius, dechanneling length
and inverse value of multiple scattering angle are proportional or
nearly proportional to the particle energy, one can widely adapt
the above estimates to various other beam energies.

The author expects that the above findings will help the experts
to estimate the perspectives of "crystal focusing" application at
future electron-positron colliders.

\begin{table}
\caption{Estimates of the $e^\pm$ beam focus sizes in both
focusing $xz$ and normal to it $yz$ planes.}
\label{tab:1}       
\begin{tabular}[b]{lllll}
\hline\noalign{\smallskip}& 1 TeV $e^-$ & 1 TeV $e^+$
&  \\
\noalign{\smallskip}\hline\noalign{\smallskip}
$\sigma_x[nm]$& 50& 20-50& \\
$\sigma_y[nm]$& 15 & 1& \\
channeling efficiency, \%& 50-70 & 70-90 & \\
\noalign{\smallskip}\hline
\end{tabular}
\end{table}
\newpage

\hspace{5 cm} R E F E R E N C E S

1. E. N. Tsyganov. "Some aspects of the mechanism of a charge
particle penetration through a monocrystal". Fermilab TM-682,
1976.

2. V. M. Biryukov, Yu. A. Chesnokov, and V. I. Kotov. Crystal
channeling and its application at high-energy accelerators.
(Springer, Berlin, Germany, 1997).

3. W. Scandale. Use of crystals for beam deflection in particle
accelerators. Mod. Phys. Lett. A. 27 (2012) 1230007.

4. V.V. Tikhomirov, V.V. Haurylavets, A.S. Lobko, and V.A.
Mechinsky. Oriented crystal applications in high energy physics.
engineering of scintillation materials and radiation technologies
(Proceedings of ISMART 2016). Springer Proceedings in Physics
Volume 200.

5. V.A. Andreev et al. Spatial focusing of 1 GeV protons by a
curved single crystal. Pis'ma Zh. Eksp. Teor. Fiz. 41 (1985) 408
[JETP Lett. 41 (1985) 500].

6. A.S. Denisov et al. First results from a study of a 70 Gev
proton beam being focused by a bent crystal. Nucl. Instr. and
Meth. B 69 (1992) 382.

7. W. Scandale et al. Observation of focusing of 400 GeV proton
beam with the help of bent crystals. Phys. Lett. B 733 (2014) 366.

8. A.G. Afonin et al. Focusing of a high-energy particle beam at
an extremely short distance. Pis'ma Zh. Eksp. Teor. Fiz. 105
(2017) 727 [JETP Lett. 105 (2017) 763].

9.  V. G. Baryshevskii, V. V. Tikhomirov. Synchrotron-type
radiation processes in crystals and polarization phenomena
accompanying them. Sov. Phys. Usp. 32 (1989) 1013.

10. V.V. Tikhomirov.  A technique to improve crystal channeling
efficiency of charged particles. JINST 2 P08006, DOI:
10.1088/1748-0221/2/08/P08006.

11. V. Guidi, A. Mazzolari and V.V. Tikhomirov. Increase in
probability of ion capture into the planar channelling regime by a
buried oxide layer. J. Phys. D: Appl. Phys. 42 (2009) 165301 DOI:
10.1088/0022-3727/42/16/165301.

12. A. Mazzolari et al. Steering of a sub-GeV electron beam
through planar channeling enhanced by rechanneling. Phys. Rev.
Lett. 112 (2014) 135503.

13. A.I. Sytov et al.  Steering of Sub-GeV electrons by ultrashort
Si and Ge bent crystals. Eur. Phys. J. C. 77 (2017) 901, DOI:
10.1140/epjc/s10052-017-5456-7.

14. V. V. Tikhomirov. Quantitative theory of channeling particle
diffusion in transverse energy in the presence of nuclear
scattering and direct evaluation of dechanneling length. Eur.
Phys. J. C 77 (2017) 483, DOI : 10.1140/epjc/s10052-017-5060-x.

15. A. Kostyuk. Crystalline undulator with a small amplitude and a
short period. Phys. Rev. Lett. 110 (2013) 115503.

16. T. N. Wistisen et al. Experimental realization of a new type
of crystalline undulator. Phys. Rev. Lett. 112 (2014) 254801.

17. V. V. Tikhomirov. Simulation of multi-GeV electron energy
losses in crystals. Nucl. Instrum. and Meth. B36 (1989) 282.

18. V.G. Baryshevsky, V.V. Tikhomirov. Crystal undulators: from
the prediction to the mature simulations.  Nucl. Instrum. and
Methods. B. 309 (2013) 30. DOI: 10.1016/j.nimb.2013.03.013.

19. A.V.Korol, A.V. Solov'yov, W. Greiner. Channeling and
Radiation in Periodically Bent Crystals, Springer Series on
Atomic, Optical, and Plasma Physics 69. Springer-Verlag Berlin
Heidelberg 2013. DOI : 10.1007/978-3-642-31895-5\_6.

 20. M. Tanabashi et al. (Particle Data Group). The Review of
Particle Physics (2018). Phys. Rev. D 98 (2018) 030001.

21. V. Guidi, L. Bandiera, V.V. Tikhomirov. Radiation generated by
single and multiple volume reflection of ultrarelativistic
electrons and positrons in bent crystals. Phys. Rev.A. 86 (2012)
042903. DOI: 10.1103/PhysRevA.86.042903.

22. L. Bandiera et al. Broad and intense radiation accompanying
multiple volume reflection of ultrarelativistic electrons in a
bent crystal. Phys. Rev. Lett. 111 (2013) 255502.  DOI: 10.1103/
PhysRevLett.111.255502.

23. L. Bandiera et al. Strong reduction of the effective radiation
length in an axially oriented scintillator crystal. Phys. Rev.
Lett. 121 (2018) 021603. DOI: 10.1103/PhysRevLett.121. 021603.

\end{document}